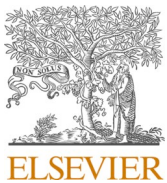
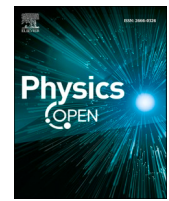

# Scalable surface ion trap design for magnetic quantum sensing and gradiometry

Qirat Iqbal, Altaf Hussain Nizamani[*]

*Institute of Physics, University of Sindh, Jamshoro, Pakistan*

A B S T R A C T

Magnetic quantum sensors based on trapped ions utilize properties of quantum mechanics which have optimized precision and beat current limits in sensor technology. Trapped ions are highly sensitive in a large span of signal ranging from DC or static B-field to the radiofrequency range in 100s of MHz and can attain the sensitivity in the range of $pT/\sqrt{Hz}$ to sub-$pT/\sqrt{Hz}$. They are tuneable to frequencies of interest and can be used as a lock-in frequency detector. This modelling and simulation based study presents an innovative design of Surface Paul Traps, enabling the use of trapped ions as ultra-sensitive sensors for magnetic field detection and precise measurement of magnetic field gradients at a sub-millimeter spatial resolution. The novel design features multiple trapping regions, allowing for the mapping of magnetic fields across various ion-trapping zones. The study demonstrates groundbreaking advancements in ion manipulation and confinement through innovative chip architecture.

## 1. Introduction

Can quantum physics be practically applied to harness its most counterintuitive properties? Physicists have been intrigued by this field ever since the formulation of quantum theory in the early 1900s. Currently, quantum computing and information and quantum cryptography are recognized as highly promising areas [1,2]. In recent years, a new class of applications has emerged, utilizing quantum mechanical systems as sensors for various physical quantities such as electric and magnetic fields, frequency and time, rotations, pressure, and biomedical applications [3]. Quantum sensors exploit the inherent sensitivity of quantum systems to external disturbances, capitalizing on their fundamental vulnerability. Despite being a relatively young field within the realm of quantum science and engineering, quantum sensing has gained extensive recognition within the physics community, resulting from decades of advancements in high-resolution spectroscopy, particularly in atomic physics and magnetic resonance. Noteworthy examples include superconducting quantum interference devices [4], atomic vapor magnetometers [5], nitrogen-vacancy (NV) color centers [6] and atomic clocks [7].

This study demonstrates the utilization of trapped ions as a precise magnetic field gradient measurement tool using novel Radio Frequency (RF) Paul Trap geometries, especially in planner trap designs. Through strategic ion trap chip design, we can trap ions in multiple regions, facilitating magnetic field mapping with millimeter to sub-millimeter spatial resolution. The proposed scalable trap design, in this study, featuring adjustable electric potential, allows trapping of number of ions above its surface and controlled transportation of these ions in linear and vertical directions. This enables the trapped ions to function as sensitive magnetic sensors, detecting B-fields at approximately $1-100 \text{ pT}/\sqrt{Hz}$ across different frequencies. By mapping magnetic fields in multiple zones, the system operates as a high-resolution magnetic gradiometer with potential for sub-milli-Tesla/mm gradient sensitivity. The ability to extend the trapping regions into two dimensions provides additional improvements to its performance. Before investigating into the new trap geometry, we will first provide a brief overview of quantum magnetometry, and then focusing specifically on the use of trapped ions.

## 2. State-of-the-art quantum magnetometry

Physical measurements with high precision play a critical role in modern science, technology, and research endeavors. Advancements in this field not only drive scientific discoveries but also provide practical tools for real-world applications and innovative research projects. Amongst the various applications, magnetic field sensing has garnered significant attention, finding utility in areas such as scanning probe microscopy, geo-magnetics, non-destructive materials evaluation, electrical measurements, and medical imaging [3]. Furthermore, magnetic field sensors contribute to the advancement of fundamental research in spin dynamics, magnetism, and mechanical motion studies. To enable direct comparisons between different quantum sensors, we employ the metric $\delta B_{min}/\sqrt{Hz}$, representing the smallest observable variation in magnetic field strength. This parameter is widely used in the literature and proves particularly useful for comparing B-filed sensors. It is important to note that $\delta B_{min}$ may vary depending on the frequency range, as the majority of these quantum sensors operate in the DC to low RF range (∼ $Hz$). In Table 1, we provide an overview of state-of-the-art magnetometers, comparing their sensitivity, frequency range, and operational limitations, including the requirement for magnetic shielding, as well as other relevant factors.

## 3. Gradiometery

Conventional magnetic sensing techniques rely on large-scale magneto-resistance or fluxgate sensors [16]. Alternatively, for measuring magnetic field gradients, Superconducting Quantum Interference Device (SQUID) detector arrays or pairs are commonly employed [4]. Similarly, diamond's nitrogen-vacancy (NV) color centers have emerged as magnetic field sensors, achieving a gradient sensitivity limit of approximately $10^{-6}$ T/nm [17]. However, these devices typically range in size from a few millimeters to approximately 100 μm with significant loss of sensitivity. In contrast to conventional approaches, trapped ions offer the potential for significantly improved magnetic gradient field measurements [18]. For instance, the Schmidt-Kaler technique exploits entangled ions in a quantum state that exclusively produces a signal in response to magnetic field gradients and the ions' extended coherence time, enabling enhanced sensitivity [19].

This study is specifically focused on designing scalable surface ion traps for ultra-high sensitive measurement of gradient magnetic fields.

## 4. Trapped ions as magnetic sensors

Trapped ions or atoms can serve as sensors due to their sensitivity to magnetic fields, which modify their energy levels. By measuring transition frequencies or shifts, the magnetic field can be determined. The concept and demonstration of using trapped ions to sense magnetic fields (B fields) has been demonstrated in various studies [20,21]. In particular, Baumgart et, al [20]. showed that for effective sensing, the frequency of the signal to be detected should match the resonance frequency between two distinct atomic states. Trapped ion magnetometers utilize tunable dressed state qubits, which offer advantages over qubits formed by bare atomic Zeeman states.

To clarify our approach for detecting magnetic fields, we use Ytterbium-171 ($^{171}$Yb$^+$) ions as the model in our methodology. It's important to understand that the sensing parameters we outline are tailored specifically for $^{171}$Yb$^+$ ions. Consequently, there could be considerable differences in these parameters when applied to different types of ions. However, the parameters defining the trap geometry in the following sections are notably independent of the ion species employed.

In a typical Paul trap arrangement, neutral $^{171}$Yb$^+$ atoms are generated from an atomic oven and channeled into the trap area. Here, the atoms are ionized and subsequently confined by employing AC and DC voltages [22]. These trapped ions can serve as magnetic field detectors. The process begins with preparing the ion in a desired quantum state tailored for the intended sensing frequency, with an auxiliary static magnetic field to fine-tune the state. Upon the ion's initialization, its ultimate state is deduced using a 369-nm laser. The presence of fluorescence at this wavelength indicates a transition to a certain state, while its absence signals an alternate state. Repeated experiments, maintaining consistent exposure time, allow for the calculation of the ion's likelihood to occupy either state. Charting these probabilities against time results in a sinusoidal graph, known as the Rabi frequency, which correlates directly with the intensity of the RF field prompting the ion state transitions [20]. Variability in the surrounding magnetic field can cause dephasing, curtailing the coherence time ($T_2$). To prolong coherence, microwave and RF dressed states are introduced by administering dual microwave fields to the ion, resulting in tunable dressed states. These states offer enhanced coherence duration and guard against magnetic disturbances at various frequencies, addressing the innate susceptibilities of bare atomic Zeeman state qubits [23].

### 4.1. Sensitivity

Trapped ion quantum magnetic sensors represent an exceptionally high-sensitivity magnetometer capable of detecting magnetic fields originating from RF and microwaves. Notably, these sensors exhibit superior sensitivity within a wide frequency range, surpassing the capabilities of other existing quantum sensors in the market. In contrast to alternative options, this sensor based on trapped ions, demonstrates remarkable sensitivity to RF and microwave radiation even at frequencies such as, in case of $^{171}$Yb$^+$; 12.6 GHz. Furthermore, it offers rapid frequency tuning capabilities, facilitated by the microwave decoupling technique employed. An advantageous aspect of this sensor is its immunity to variations in the external magnetic field, obviating the need for cumbersome shielding or a controlled laboratory environment for its operation. The sensitivity a magnetometer based on this method can achieve is defined as [20],

$$|\delta B_s|_{min} = \frac{2.33\hbar}{\mu_B\sqrt{nT_{tot}T_2}} \quad (1)$$

with '$n$' the number of ions, $T_2$ is the coherence time and $T_{tot}$ is total sensing time. Whereas, the sensitivity of a magnetometer using entangled ions can be calculated by:

$$|\delta B_{s,entangled}|_{min} = \frac{|\delta B_s|_{min}}{\sqrt{n}} \quad (2)$$

A single trapped atomic ion has been demonstrated as a magnetic sensor with a 4 pT/$\sqrt{Hz}$ sensitivity for alternating-current (RF) magnetic fields up to MHz range [20]. This sensitivity can be enhanced by increasing the number of trapped ions, since above mentioned principles (formulas) shows;

$$Sensitivity \propto \frac{1}{\sqrt{n}} \quad (3)$$

and for entangled ions above equation reduces to,

**Table 1**
Comparisons of state-of-the-art magnetometers.

| Sensor | Type | Sensitivity (Hz$^{1/2}$) | Frequency Range | Shielding |
| --- | --- | --- | --- | --- |
| SERF [8] | Vapor Cell | 1pT to 160aT | 40Hz–200Hz | Yes |
| Search coil [9] | Induction coil | ∼100fT | ∼650kHz | No |
| SQUID [10] | Superconductor | 0.08fT | 423kHz | Yes |
| Artificial Atom [11] | Super Conducting PCQ | ∼3pT | 10MHz | Yes |
| Flux-gate [12] | Coils | ∼200pT | 100Hz | No |
| Alkali metal [5] | Vapor Cell | ∼1fT | 100kHz | Yes |
| BEC-Ensemble [13] | BEC | ∼10pT | – | Yes |
| NV Centers [14] | Diamond | ∼4nT | 1.5MHz | No |
| Tunable AM [15] | Vapor Cell | ∼2fT | ∼500kHz | Yes |

$$Sensitivity_{(entangled\ ions)} \propto \frac{1}{n} \quad (4)$$

This shows that trapping of entangled ions can enhance the sensitivity (in DC field) of the sensor twice that of untangled ions [24].

### 4.2. Tunability

Fundamentally, trapped ion quantum sensor detects oscillating magnetic waves by observing how rapidly they can drive a transition between two arbitrary quantum states. The work of Baumgart et al. [20] served as the basis for the sensing method summaries here. The two states which are used in sensing are indicated by; $|\downarrow\rangle \equiv (0,1)^T$ and $|\uparrow\rangle \equiv (1,0)^T$ having the zero-point energy positioned halfway between these two states. Multiple transitions are seen on the $^2S_{1/2}$ manifold in Fig. 1, which may be used to monitor magnetic fields by determining their Rabi frequencies in the manner outlined in previous section. This comprises the $|0'\rangle \leftrightarrow |+1\rangle$ & $|-1\rangle$ transition for detecting radio frequency signal (1 MHz–150 MHz) and the $|0\rangle \leftrightarrow |\pm 1\rangle$ transition for sensing microwaves (12.5 GHz–12.7 GHz). This method, known as "bare" state sensing, operates without the use of dressing fields and is highly sensitive to the slightest changes in the magnetic field due to its reliance on hyperfine states. Variations in the magnetic field alter the splitting between these states, leading to changes in the resonant frequency and causing the quantum superposition of the states to become decoherent. This decoherence shortens the possible duration of experiments and reduces their overall sensitivity. To counteract these limitations, the "dressed states" technique is employed for more stable detection of RF and microwave signals. This method enhances the stability of the sensor, allowing for longer coherence times and improved sensitivity. The method is described in Ref. [25]. Further to this, we can regulate the frequency splitting by using the hyperfine states $\omega+$, $\omega-$ and $\omega_o$. Hence, by adjusting the applied B-field, the system may be tuned to any desired frequency to be detected. The steps required in sensing with microwaves and RF are fairly similar, with the exception that we can utilize the $|0'\rangle$ as an intermediate state between both the $|+1\rangle$ and $|-1\rangle$ rather than the $|0\rangle$. As seen in Fig. 1, this is achieved by driving the $|0'\rangle \leftrightarrow |-1\rangle$ and $|0'\rangle \leftrightarrow |+1\rangle$, transitions with an applied RF-field. In this way the trapped ion magnetometer can be tuned to sense RF signal in particular range or specific frequency and could work as tunable magnetometer.

## 5. Scalable surfac ion tap designs

The trap design outlined in this study is founded on the principles of a Paul trap, wherein ions are confined through the application of combined AC and DC voltages. This study is limited to simulation and modelling of trap designs. Upon finalizing the design, these trap chips can be efficiently produced utilizing existing microfabrication techniques [26,27]. For achieving ultra-high sensitivity in magnetic field sensing and field gradient measurements, it is essential to carefully engineer an ion trap chip capable of confining a substantial number of ions across multiple regions.

### 5.1. Trap modelling

Using the analytic method as described by M. G. House [28], we have simulated our surface ion trap design in Mathematica software. A trap made of only RF-electrodes would confine an ion in radial directions only but it is free to move axially. In order to control the trapped ions in axial direction the DC-electrodes are also needed as shown in Fig. 2.

Once the electrostatic field generated by the trap electrodes is modeled, the RF trapping potential (pseudo-potential approximation) created by the RF-electrodes above the surface of the trap can be calculated using analytic method with gapless approximation using equation below;

$$\psi(x,y,z) = \frac{Z^2 e^2}{4m\Omega^2} |\nabla \varphi_{rf}|^2 \quad (5)$$

where $\Omega$ is drive RF-frequency, $m$ is mass of ion, $Z$ is charge-number and $\varphi_{rf}$ is electric field generated by the applied AC voltage with a drive frequency $\Omega_{rf}$.

Fig. 2 demonstrates the trapping potential created above the surface of a flat ion trap configuration. In this illustration, the trapping region, referred to as RF-nil, ideally should possess a potential of zero where the ions can be trapped.

### 5.2. Trap parameters

#### 5.2.1. Ion height

The ion height position 'h' shown in Fig. 2, measures the vertical separation between the trap's surface and the RF nil point. The ion height position 'h' of the trapping regions can be calculated using geometrical factors (electrode widths) of the traps; 'a', 'b' and 'c' as shown in Fig. 2 by using the following equation;

$$h = \frac{\sqrt{abc(a+b+c)}}{b+c} \quad (6)$$

where 'b' and 'c' are widths of RF electrodes and 'a' is width of ground electrode in between RF electrodes.

Reducing the height of trapped ions results in increased rates of motional heating, while increasing their height leads to a weakened trapping potential.

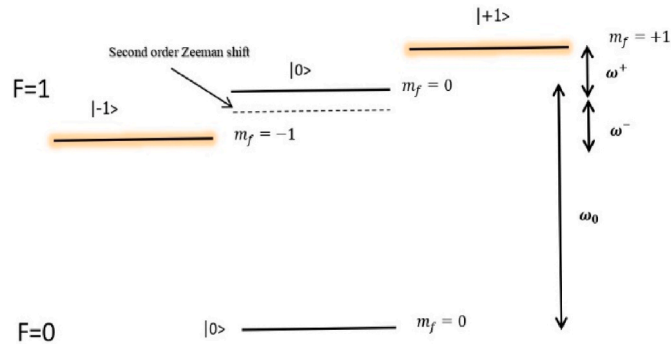

**Fig. 1.** The $^{171}$Yb$^+$ hyperfine $^2S_{1/2}$ manifold, that contains four states; the $|F=0\rangle$ ground state $|0\rangle$; the $|F=1\rangle$; $m_f = 0$ state denoted by $|0'\rangle$; and two hyperfine splitting $m_f = +1$ and $m_f = -1$, which are denoted $|+1\rangle$ and $|-1\rangle$.

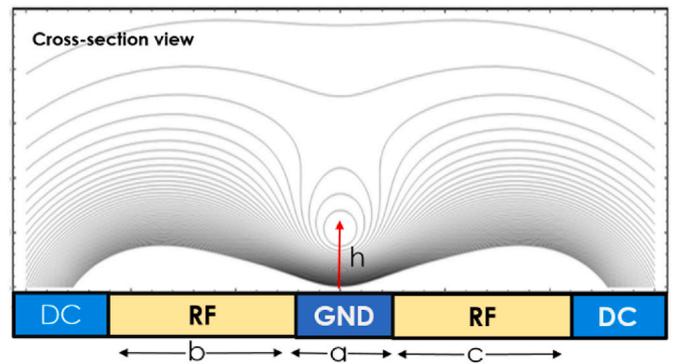

**Fig. 2.** Cross-sectional representation of the trap potential, also referred to as pseudo-potential, demonstrates the contour plot above the surface area of a singular trap region. This illustration includes the positioning of the ground electrode with a width labeled 'a,' alongside RF electrodes with widths 'b' and 'c' being equal. Additionally, the ion height above the surface is denoted as 'h.' The potential within this setup is generated through the application of AC voltage ($V_{rf}$) to the RF electrodes. The standard electrode widths typically fall within the range of a few hundred micrometers.

### 5.2.2. Trap depth

The lowest amount of energy needed for an ion to escape the trap is known as the trap depth. Also, the distinction between the escaping point and the trap potential at the local minimum (RF-Nil), which follows the identification of the escape point, is referred to as the trap depth. Trap depth could be calculated using the equation;

$$T.D = \frac{\Gamma}{h^2}\kappa \qquad (7)$$

where,

$$\Gamma = \frac{e^2 V_{rf}^2}{\pi^2 m \Omega^2} \qquad (8)$$

Here '$\kappa$' is the geometric factor and can be optimized for achieving higher trap depth [29].

### 5.2.3. Secular frequency

Secular frequency can be defined as the characteristic frequency of oscillation displayed by trapped ions within a harmonic potential. It represents the frequency at which the ions oscillate around stable equilibrium positions in the trap, influenced by the trapping potential and the ions' mass. The secular frequency is a crucial parameter in ion trapping experiments, impacting the ions' stability, sensitivity, and the feasibility of performing quantum operations. Secular frequency $\omega_s$ at a certain ion height $h$ can be calculated by given equation;

$$\omega_s = \frac{eV_{rf}}{\sqrt{2}\, mh^2\Omega} \qquad (9)$$

## 5.3. Scalable trap geometry

The trap design discussed here is made scalable by creating multiple trapping regions where a number of ions can be trapped. We can achieve confinement of individual or number of ions in multiple regions within a linear ion trap on the chip by applying a combination of static and oscillating electric potentials to integrated electrodes as shown in Fig. 3. The electrodes in the trap are configured to create regions of electric potential at various locations using a DC field, fulfilling two key functions. First, they can trap ions selectively in certain areas. Second, they allow for the movement of these ions between areas using shuttling protocols. Isolating ions within these regions ensures they operate independently, maintaining the information encoded in their states for later use. An integrated ion-trap chip is the preferred platform for creating a scalable system based on these principles.

These traps can be fabricated through lithographic patterning from a monolithic semiconductor substrate, eliminating the need for manual assembly and alignment of individual electrodes. Given the capabilities of current fabrication technology, it appears feasible to scale this structure to accommodate hundreds or even thousands of electrodes and various on chip features [30–34],

To develop a scalable trap, initially we designed a trap with two RF electrodes and six DC electrodes, creating a single trapping region. To expand this into multiple trapping zones, we replicated this design. This resulted in a setup with four long RF electrodes separated by ground electrodes, and twenty four DC electrodes, capable of forming four distinct trapping regions as shown in Fig. 3(a) and (b). This methodology allows us to enhance the trap geometry, facilitating the creation of multiple trapping regions in a similar manner. This setup is crucial for enabling two-dimensional magnetic field mapping for gradiometry. Fig. 4 illustrates the 3D model view of multi-regional trap with approximate geometric parameters.

## 6. Trap optimisation

### 6.1. Optimisation of trap parameters

#### 6.1.1. Geometrical parameters

The surface trap design in this study has been fine-tuned to achieve the maximum trap depth at a specific ion height. To accomplish this, the geometric parameters shown in Fig. 2 must be optimized accordingly. Here, we have optimized the sensor trap using approach given by Nizamani et, al [29]. Initially, trap geometric parameters such as ion height (ion-electrode distance) '$h$' for the trap needs to be decided. In real-world ion trap experiments, for better laser access, and lower the heating rate, keeping $h > 120\mu m$ is an appropriate choice. The heating rate increases rapidly when the ion-electrode distance reduced and proportion to $h^{-4}$ [35]. The widths '$b$' and '$c$' of the RF-electrodes for the mentioned ion height '$h$' above the surface of the trap should be $\sim 315\mu m$, and the separation between the RF electrodes (here, it is the width '$a$' of the central ground/control electrode) should be $\sim 85\mu m$ to obtain the maximum trap depth by applying an AC-voltage with a RF-drive frequency $\Omega_{rf}$. For this configuration, the segmented DC electrode's widths should be $\sim 310\mu m$ while they are functioning as control electrodes in axial direction. By using DC electrodes of these sizes and adjusting the applied static voltages on these electrodes to relax the trap potential in the axial direction, it becomes feasible to accommodate a substantial number of ions within the trapping regions. This capability is of great significance as it serves to enhance the sensitivity of an ion trap magnetometer. Further to this by varying the DC voltages, linear transportation of the trapped ions can also be achieved [36].

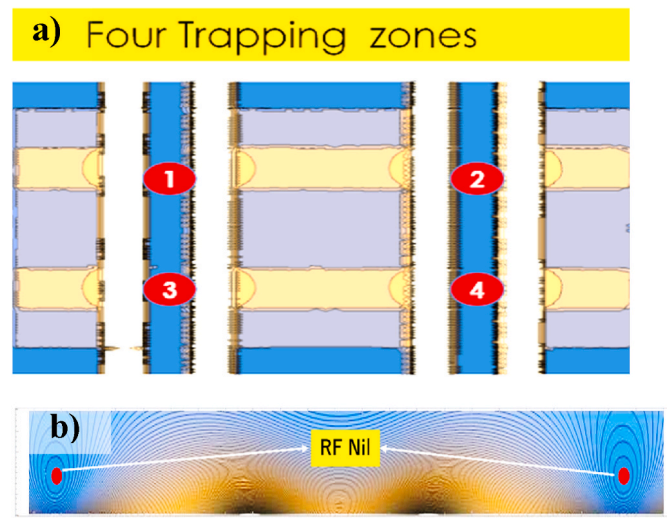

**Fig. 3.** (a) Surface Paul trap design for four trapping regions. The long rails are RF-electrodes and side short rails are modeled as DC (control) electrodes. Whereas, four trapping zones, above the surface of the trap, are highlighted in red. (b) Cross section view of trap potential at one of trapping zones above the surface of the trap. (For interpretation of the references to color in this figure legend, the reader is referred to the Web version of this article.)

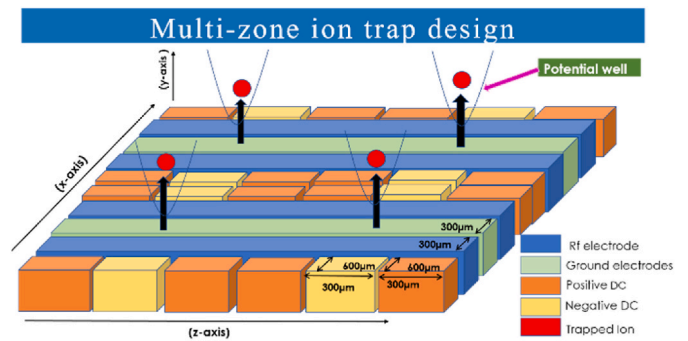

**Fig. 4.** Illustrating a Multi-Zone Ion Trap Design in Three Dimensions: Demonstrating the trap's approximate length along the $x$ and $z$ dimensions, while the trapped ions are positioned in the $y$ dimension above the trap's surface.

### 6.2. Voltages

To determine the ideal operational AC voltage ($V_{rf}$) to be applied on the trap electrodes that will yield substantial trap depth and the desired secular frequency at the ion's height position within a trapping region, certain factors must be considered. In practical experiments, one limiting factor is the breakdown voltage of the trap electrodes, which imposes a maximum limit on the applied voltage. Another crucial limitation stems from power dissipation within the trap, potentially leading to trap heating and increased anomalous heating of the trapped ion, as well as the outgassing from the trap material. These factors collectively influence the permissible level of RF voltage that can be applied on the RF-electrodes in a Paul-trap. The breakdown voltage depends on trap insulator materials, on which electrodes are fabricated, the spacing between the trap electrodes, and the level of vacuum within chamber where the trap is mounted. An AC ($V_{RF}$) voltage up to 500 $V_{ac}$ can be applied safely for a sensor trap chip with the gap between the electrodes above 5μm as discussed in Ref. [37]. All the optimized parameters including geometric, and voltages are summarized in Table 2.

### 6.3. Impact of scalable multi-rail trap design

#### 6.3.1. Impact on trap parameters

When AC and DC operating voltages are introduced to the electrodes of the trap, several potential wells emerge across various areas in the multi-rail RF-trap configuration. These wells exhibit subtle differences compared to a single RF-trap design, especially as voltages are activated in separate trapping areas or collectively across all regions, as depicted in Fig. 4. The effects of the electric potential created by the multi-rail RF and DC electrodes on trap characteristics, such as trap depth and secular frequencies in the *x*, *y*, and *z* directions, are compiled and presented in Table 3. Notably, activating all electrodes simultaneously leads to a significant increase in ion height. In contrast, the secular frequencies exhibit a slight decrease across all directions (*x*, *y*, and *z*), as detailed in Table 3.

#### 6.3.2. Rotation of principle axis

In practical surface ion trap experiments, the cooling laser beams are directed parallel to the trap surface but are intentionally kept from making direct contact with it. Consequently, they do not effectively cool the ion motion along the primary axis aligned with the surface in the axial direction. This challenge can generally be addressed by altering the orientation of the principal axis of RF-nil. The process of rotating can be accomplished by adjusting the overall electric potential at the location of the ion. This is commonly done by employing an uneven configuration of RF-electrodes, in which one electrode is broader than the other. Alternatively, a segmented ground electrode can be placed under the ion, to which a combination of positive and negative DC voltage is applied [31]. Nonetheless, these methods result in a notable decrease in the depth of the trap. In our multi-region trap design, we have structured each RF electrode to generate an electric potential that influences neighboring trapping regions. As a result, the principal axis naturally undergoes rotation by default, without the need to disrupt the symmetry of the RF electrodes. This rotation is a favorable outcome of our trap design. It's worth noting that without this approach, altering the symmetry of the RF-electrodes would have had a significant impact on trap parameters, especially trap depth. As shown in Fig. 5, the principal axis has been successfully rotated by an angle of at least 6°, which is sufficient for laser cooling of the trapped ions above the surface.

**Table 2**
Summary optimized voltages and geometric trap parameters.

| Optimized geometric and voltage parameters | |
|---|---|
| Width of ground electrode (a) | 85μm |
| Width of RF electrode (b = c) | 315μm |
| Width of DC electrode | 310μm |
| Voltage on RF electrode ($V_{RF}$) | 200V |
| Voltage on + ve DC electrode | 6V |
| Voltage on -ve DC electrode | −8.4V |
| Drive Frequency(Ω) | 2π x 22MHz |

**Table 3**
A comparison of various trap parameters influenced by applied voltages.

| Parameter: | Two RF electrodes ON | Four RF electrodes ON | All RF& DC electrodes ON |
|---|---|---|---|
| Ion height (μm) | 123 | 136 | 133 |
| Trap depth (eV) | 0.177 | 0.112 | 0.16 |
| X-secular Freq. (MHz) | 1.81 | 1.4 | 1.4 |
| Y-secular Freq. (MHz) | 1.92 | 1.5 | 1.5 |
| Z-secular Freq. (MHz) | – | – | 0.2 |

### 7. Conclusion

We have successfully designed an innovative and optimized surface ion trap chip design that exhibits the capability to trap ions within different trapping zones, enabling us to harness their unique properties for RF-magnetic field sensing. In these distinct trapping zones, the trapped ions can function as quantum sensors with exceptional sensitivity, ranging from approximately 1 to 100 pT/$\sqrt{Hz}$ across various frequency ranges. Taking our capabilities a step further, we can utilize trapped ions in these diverse zones to create a magnetic gradiometer with unprecedent level of sensitivity. By precisely mapping the magnetic field around these trapped ions in their respective zones, we enable them to operate as highly sensitive devices capable of detecting subtle variations in magnetic field strength. The spatial resolution achieved by these ion-based magnetic gradiometers spans from sub-millimeter to millimeter levels. This means that we can effectively measure magnetic field gradients in the order of pT/mm.

To extend spatial resolution capabilities in three dimensions, we can implement a technique involving the vertical movement of trapped ions within the various trapping zones. This vertical transportation can be achieved by applying DC voltage to the ground/control electrodes (in between RF-electrodes) beneath the trapped ions. As a result, we can further enhance our ability to measure magnetic field gradients with extraordinary precision in all spatial dimensions.

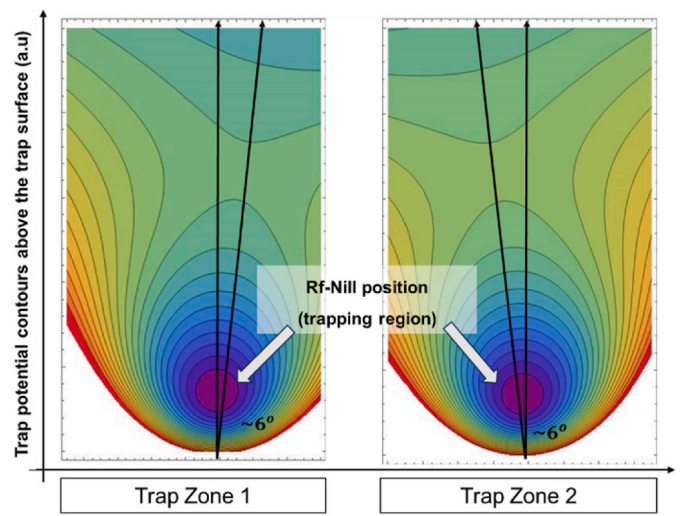

**Fig. 5.** Approximately 6° rotation of principal axis at the trapping regions due to mutual impact of RF electric field of multi-rails trap geometry.

**Future work**

In a surface trap design, the ability to detect a magnetic field in two dimensions can be achieved by confining ions within different trapping regions. We have developed a microchip design that incorporates multiple trapping zones, enabling the simultaneous trapping of ions in these zones. Consequently, this chip possesses the capability to sense magnetic fields in two dimensions. Furthermore, our research offers the potential for expanding these trap designs to include the trapping of ions in two-dimensional arrays, as well as facilitating the vertical and linear transport of trapped ion. Consequently, these designs possess the capacity to map magnetic fields in all three dimensions, thereby functioning as a 3D-magnetic gradiometer.

**CRediT authorship contribution statement**

**Qirat Iqbal:** Writing – original draft. **Altaf Hussain Nizamani:** Supervision.

**Declaration of competing interest**

The authors declare the following financial interests/personal relationships which may be considered as potential competing interests:
Altaf Hussain Nizamani reports was provided by University of Sindh. Altaf Hussain Nizamani reports a relationship with University of Sindh that includes: employment.

**Data availability**

No data was used for the research described in the article.